\documentstyle[twoside,fleqn,espcrc2,epsf]{article}

\def\spose#1{\hbox to 0pt{#1\hss}}
\def\ltapprox{\mathrel{\spose{\lower 3pt\hbox{$\mathchar"218$}}
 \raise 2.0pt\hbox{$\mathchar"13C$}}}
\def\gtapprox{\mathrel{\spose{\lower 3pt\hbox{$\mathchar"218$}}
 \raise 2.0pt\hbox{$\mathchar"13E$}}}
\def\inapprox{\mathrel{\spose{\lower 3pt\hbox{$\mathchar"218$}}
 \raise 2.0pt\hbox{$\mathchar"232$}}}

\newcommand{\id}{\mbox{1$\!\!$I}}

\renewcommand{\Pr}{\hat{\mbox{I}\!\!\mbox{P}}}

\newcommand{\bfn}{\mbox{\protect\boldmath{$n$}}}
\newcommand{\bfns}{\mbox{\protect\boldmath{$\scriptstyle n$}}}
\newcommand{\bfnps}{\mbox{\protect\boldmath{$\scriptstyle{n'}$}}}
\newcommand{\bfnp}{\mbox{\protect\boldmath{$n'$}}}
\newcommand{\bfeigs}{\mbox{\protect\boldmath{$\scriptstyle{8,1}$}}}
\newcommand{\bfeig}{\mbox{\protect\boldmath{$8,1$}}}

\newcommand{\<}{\langle}
\renewcommand{\>}{\rangle}
\newcommand{\beq}{\begin{equation}}
\newcommand{\eeq}{\end{equation}}
\newcommand{\beqn}{\begin{eqnarray}}
\newcommand{\eeqn}{\end{eqnarray}}

\title{Non-Perturbative Renormalization of Lattice QCD.}

\author{G.C.~Rossi
\address{Dip. di Fisica, Universit\`a di Roma ``Tor Vergata''
and I.N.F.N., Sezione di Roma II, \\ 
Via della Ricerca Scientifica 1, I-00133 Roma - Italy}%
\thanks{Talk presented at LATTICE96 (theoretical develop-ments)}}

\begin{document}

\begin{abstract}
In this talk I will discuss a number of approaches designed to deal with
the problem of setting up a fully non-perturbative renormalization
procedure in lattice QCD. Methods based on Ward-Takahashi identities 
on hadronic states, on imposing chiral selection rules on
amplitudes with external quark/gluon legs, on
the use of the Schr\"{o}dinger functional and on ``heating and cooling''
Monte Carlo steps are reviewed. I conclude with some remarks on the
possibility of defining next order terms (higher twists,
``condensates", ...) in short distance expansions.

\end{abstract}

\maketitle

\section{Introduction}
Renormalization is a necessary step to bring numbers extracted from Monte
Carlo simulations in contact with actual physical data. Its role is
threefold:

$\bullet$ to allow the construction of finite operators;

$\bullet$ to recover (modulo possible anomalies) chiral symmetry, broken by 
lattice regularization;

$\bullet$ to connect the high momentum perturbative regime of QCD with the 
low momentum non-perturbative region of the theory through the running of
renormalized quantities.

Perturbative calculations of Renormalization Constants and Mixing
Coefficients (RC\&MC's) can only be of limited value in numerical
simulations, because

$\bullet$ uncontrollable Non-Perturbative (NP) contributions may affect 
dimensional MC's,

$\bullet$ only the lowest order terms of perturbative expansions can be
calculated,

$\bullet$ field theoretical perturbative series are (most probably) only
asymptotic.

Methods to compute RC\&MC's beyond Perturbation Theory (PT) are thus 
highly desirable, if not for the fact that, in a strict sense, in a 
fully NP approach of QCD use of PT should nowhere be made.
In the recent past, and especially in the last year, a number of
significant progresses have been made in the direction of developing
efficient strategies to compute RC\&MC's in a NP way, i.e.~directly
from appropriately designed Monte Carlo simulations. 
The plan of the talk is the following. In Sect.~2 I start by recalling 
how Ward-Takahashi identities (WTI's) on hadronic states can be used 
i) to define partially conserved 
$SU(N_f)_L\otimes SU(N_f)_R$ currents, obeying Current Algebra (CA) and 
ii) to construct finite composite operators with well defined chiral 
transformation properties. In Sect.~3 I discuss the NP determination of 
RC\&MC's, based on imposing chiral selection rules and renormalization 
conditions on Green functions with (amputated) quark/gluon external 
legs. This approach amounts to require the validity of WTI's on quark/gluon 
states. In Sect.~4 the use of the Schr\"{o}dinger Functional
(SF) for the determination of the O($a$) improved form of the fermionic action
and of quark bilinear RC\&MC's is illustrated.
For completeness I will briefly recall in Sect.~5 the ``heating-cooling''
method designed to extract from pure gauge Monte Carlo simulations
the RC of the topological charge density and the divergent subtraction needed 
to arrive at a finite definition of the topological susceptibility. Finally I 
conclude in Sect.~6 with a few observations on the problem of giving a 
rigorous NP definition of large scale  effective expansions
beyond leading order  and on the related question of the possibility of
evaluating the ensuing higher dimensional corrections.

\section{WTI's on hadronic states}

For the purpose of this talk lattice QCD should be considered as a
regularized version of the theory described by the (euclidean) continuum
Lagrangian
\begin{equation}
\begin{array}{l}
{\cal L}_{QCD} \equiv {\cal L}_{YM} + {\cal L}_F = \frac{1}{2}
\mbox{Tr}(F_{\mu\nu}F_{\mu\nu})+ \\
+ \bar\psi(\gamma_\mu\partial_\mu +i g \gamma_\mu A_\mu +m)\psi
\label{eq:LQCD}
\end{array}
\end{equation}
As is well known, to avoid fermion species doubling a dimension 5
(irrelevant) operator, the so-called Wilson term~\cite{W}, must be added to
the naive lattice discretization of~(\ref{eq:LQCD}), leading to
the following expression for the fermionic part of the action
\[
\begin{array}{l} 
S_F^L = a^4 \sum_x
\Bigl[{-1\over 2a} \sum_\mu [\bar\psi (x) U_\mu (x) (r-\gamma_\mu) \psi
(x+\mu) \\ +\bar\psi (x+\mu) U_\mu^{\dagger} (x) (r+\gamma_\mu) \psi(x)]+
\\+\bar\psi (x)(M_0 + {4r\over a})\psi (x)\Bigr]
\end{array}
\]
The Wilson term ($r\int d^4x \bar\psi D^2 \psi$ in continuum notations) 
explicitly breaks global $SU(N_f)_L\otimes SU(N_f)_R$ chiral symmetry. 
Chiral symmetry is an exact property of the continuum Lagrangian,
which is only softly broken by quark mass terms. Actually the 
Lagrangian~(\ref{eq:LQCD}) possesses two further $U(1)$ symmetries: an
exact $U(1)_V$ vector symmetry, associated to the conservation of fermion
number, and an axial $U(1)_A$ symmetry, which is explicitly broken at quantum 
level by the color anomaly and is responsible for the singlet/non-singlet 
pseudoscalar mass splitting. The same symmetry pattern is seen to emerge from 
the $a\rightarrow 0$ limit of the regularized theory.

\subsection{Vector and axial currents}
It has been shown, in fact, in refs.~\cite{KG}~\cite{BM} that in the
continuum limit chiral symmetry can be fully recovered. Partially conserved 
vector and axial currents, $\widehat V^f_\mu$ and $\widehat A^f_\mu$, 
$f=1,...,N_f^2-1$, satisfying CA can be defined. In lattice QCD finite RC's 
relate, in the limit $a\rightarrow 0$, operators obeying CA to their bare 
expressions 
\begin{equation} 
\widehat{V}_\mu^f = Z_V V_\mu^f\,,\qquad \widehat{A}_\mu^f = Z_A A^f_\mu
\label{eq:CURRENTS} 
\end{equation}
The RC's $Z_V$ and $Z_A$ can be determined non-perturbatively precisely by
requiring the validity of the non-linear constraints coming from
CA~\cite{BM}~\cite{LMM} \footnote{Note that the expression of the bare
currents is not uniquely determined, but can be modified by the addition of
higher dimensional operators, vanishing in the limit $a\rightarrow 0$, 
with the same conserved quantum numbers as the currents. In doing so the 
RC's $Z_V$ and $Z_A$ will have to be accordingly modified. A similar 
observation holds for any lattice operator.}. Once $Z_A$ has been determined, 
imposing PCAC in the (naive) continuum form fixes the linearly divergent mass 
subtraction term, $\bar M$, which allows to make the pseudoscalar operator, 
$\bar\psi \gamma_5 \{\lambda^f,M_0-\bar M\} \psi$, finite.

\subsection{Other composite operators}
The explicit breaking of chirality, induced by the Wilson term, also spoils 
the ``nominal" chiral properties of composite lattice operators. Let
$O_{[{\bfns}]}=\{O^i_{[{\bfns}]}\}$ be a basis of operators which naively 
(i.e. at the tree level) transform according to the irreducible 
representation, $[\bfn]$, of the chiral group
\beq \frac {1}{i}{\delta O^i_{[{\bfns}]}(0)\over \delta
\alpha^f}=(r^f_{[{\bfns}]})^{ij} O^j_{[{\bfns}]}(0)
\label{eq:(7.1)}
\eeq 
where I have denoted by $\delta O^i_{[{\bfns}]}/\delta \alpha^f$ the
variation of $O^i_{[{\bfns}]}$ under an infinitesimal global chiral rotation 
and by $r^f_{[{\bfns}]}$ the $f^{th}$ generator of the chiral group in the
representation $[{\bfn}]$. 

Explicit lattice perturbative calculations~\cite{MBSD} show that, as expected,
radiative corrections induce mixing among operators  with different
(nominal) chiral properties. The existence of partially conserved lattice 
vector and axial currents guarantees, however, that, given the set of bare 
operators $O_{[{\bfns}]}$, it will be possible to determine the coefficients, 
$c^{ij}_{[{\bfns},{\bfnps}]}$, of the linear combination
\beq 
\widetilde{O}^i_{[{\bfns}]}= O^i_{[{\bfns}]}+\sum_{{\bfnps},j}
c^{ij}_{[{\bfns},{\bfnps}]}O^j_{[{\bfnps}]}
\label{eq:MIXING}
\eeq 
in such a way that, up to terms of O$(a)$, $\widetilde{O}^i_{[{\bfns}]}$ 
will obey the set of WTI's appropriate for an operator belonging to the 
representation $[{\bfn}]$. The resulting operator will automatically 
be multiplicatively renormalizable. 
In~(\ref{eq:MIXING}) the indices $[{\bfnp}]$ and $j$ run over all operators 
with dimensions equal or smaller than $O_{[{\bfns}]}$ and belonging to all 
possible representations of the chiral group with the only constraint 
of having the same conserved quantum numbers as $O_{[{\bfns}]}$. The mixing
coefficients of operators with the same dimension as $O^i_{[{\bfns}]}$ are
dimensionless and finite, while the coefficients of the lower dimensional
ones will diverge as $c^{ij}_{[{\bfns},{\bfnps}]} \sim a^{-(\mbox{dim}
O^i_{[{\bfns}]}-\mbox{dim}O^j_{[{\bfnps}]})}$, in the limit $a\rightarrow 0$.

In this talk for lack of space I will limit my considerations to the 
massless case. Since flavor vector symmetry is unbroken by the lattice 
regularization, to determine the $c$'s it will suffice to look at the
axial WTI's. One way to proceed is to impose that the 
renormalized, integrated, lattice axial WTI 
\begin{equation}
\begin{array}{l}
\sum_{x} 
\nabla_\mu \<h_1\vert T( (\widehat A^f_\mu (x)-\bar\chi_A^f(x))\widetilde 
O^i_{[{\bfns}]}(0))\,\vert h_2\>= \\
=-i\<h_1\vert {\delta\widetilde{O}_{[{\bfns}]}^i(0)\over\delta\alpha^f} 
\,\vert h_2\>
\end{array}
\end{equation}
taken between on-shell mesonic states, $\vert h_1\>$ and 
$\vert h_2\>$ \footnote{As chiral symmetry is spontaneously broken,
the axial rotations of the $O_{h_1}$ and $O_{h_2}$ operators, which create the
mesons $h_1$ and $h_2$ from the vacuum, do not
contribute to on-shell matrix elements.}, has the form expected 
for an operator belonging to the representation 
$[{\bfn}]$, i.e. that
\[
\begin{array}{l}
\sum_{x}\<h_1\vert T(\bar\chi^f_A (x) \widetilde O^i_{[{\bfns}]}
(0) )\,\vert h_2 \> - i \<h_1\vert  {\delta\widetilde O^i_{[{\bfns}]}(0)\over
\delta \alpha^f}\,\vert h_2 \> \\
= (r^f_{[{\bfns}]})^{ij} \<h_1\vert \widetilde O^j_{[{\bfns}]} (0) \,\vert h_2\>
\end{array}
\]
By varying the external states,  one
can write a sufficiently large number of (linear) equations and fix the
$c$'s uniquely. $\widetilde O_{[{\bfns}]}$ will be finally made finite
multiplying it by an overall RC defined, for instance, by the condition
\[
\begin{array}{l}
\<h_1\vert  \widehat O_{[{\bfns}]}(\mu)\vert h_2\>|_{_{\mu ^2}} \equiv
\<h_1\vert Z(a\mu ) \widetilde O_{[{\bfns}]}(a) \vert h_2\>|_{_{\mu^2}}= \\ 
= {\mbox{continuum normalization}}
\end{array}
\]
In practice this procedure can only be employed for bilinear quark
operators (such as currents, scalar and pseudoscalar densities,...). For
more complicated operators, like the four-quark operators describing the
Effective Non-Leptonic Weak Hamiltonian (ENLWH), this approach would require
measuring with practically unattainable precision a much too large number of
matrix elements \footnote{See, however, the method
proposed in~\cite{BM}, based on the idea of fixing the power divergent MC's
from the relations coming from the low energy theorems of the chiral
symmetry (Soft Pions Theorems - SPT's), while computing in PT the overall 
RC and all the other dimensionless MC's.}. In the following sections I will 
describe alternative strategies, recently proposed to overcome this kind of
difficulties.

\section{WTI's on quark and gluon states}
The simple idea of~\cite{NP} is to compute RC\&MC's mimicking in
Monte Carlo simulations the straightforward procedure employed
in continuum PT. To explain the method let me consider the physically
interesting case of the $\Delta S=2$ four-quark operator $O_{0}^{\Delta S=2}
=(\bar s \gamma_\mu^L d)(\bar s \gamma_\mu^L d)$, whose matrix elements
control the $K\rightarrow \pi\pi\vert_{\Delta I=3/2}$ decay 
amplitudes~\cite{DS}~\cite{TA}.
In lattice QCD this operator mixes with 4 other operators of dimension 6 with
dimensionless coefficients. The finite operator, $\widehat O^{\Delta S=2}$, has
the general expression
\begin{equation}
\widehat O^{\Delta S=2}=Z^{\Delta S=2}\big[O^{\Delta S=2}_{0}+\sum_{i=1}^4
Z^{\Delta S=2}_i O^{\Delta S=2}_{i} \big]
\label{eq:DeltaS}
\end{equation}
The proposal of~\cite{NP} is to extract from Monte Carlo data the 
$\<\bar s d\vert O^{\Delta S=2}_{i} \vert \bar d s\>, i=0,1,...,4$ matrix 
elements and to fix the coefficients $Z^{\Delta S=2}_i$, by requiring that,
at large $p_k^2=\mu^2$ (the $p_k$'s are the momenta of the external legs), all 
chirality violating form factors of the renormalized operator, 
$\widehat O^{\Delta S=2}$, should be identically zero. In practice this means 
that in the interacting theory $\widehat O^{\Delta S=2}$ must precisely match, 
at large $\mu^2$, the flavor, 
color, spin,... structure of the bare operator, $O_{0}^{\Delta S=2}$. 
The overall RC can be successively fixed by using, for instance, the same 
renormalization condition employed in the continuum. Technically the whole 
procedure is carried over by first defining the matrix ($i,j=0,1,...,4$)
\begin{equation}
D_{ij}(\mu) = {\mbox{Tr}}(\Pr_i 
\Lambda^{\Delta S=2}_{j}(\bar{q} q\bar{q} q;\mu)) 
\label{eq:AMPPR}
\end{equation}
where the amplitude $\Lambda^{\Delta S=2}_{j}(\bar{q} q\bar{q} q;\mu)$ is the 
four-legs amputated matrix element of the operator $O^{\Delta S=2}_{i}$ 
with each leg taken at momentum $p^2=\mu^2$
\[
\Lambda^{\Delta S=2}_{i}(\bar{q} q\bar{q} q;\mu)=\<\bar{q}(p) q(p) 
O^{\Delta S=2}_{i} q(p) \bar{q}(p)\>\vert _{p^2=\mu^2}^{Amp}
\]
The~$\Pr$'s~are~orthogonal~projectors, ${\mbox{Tr}}(\Pr_i\Pr_j)=\delta_{ij}$,
satisfying ${\mbox{Tr}}(\Pr_i \Lambda^{\Delta S=2}_{j}(\bar{q} q\bar{q}
q)\vert_{tree}) = \delta_{ij}$. MC's are determined from the constraints
\footnote{In the equations below for notational simplicity we drop the 
arguments of the amplitudes $\Lambda$ whenever unnecessary.}
\begin{equation}
\begin{array}{l}
\mbox{Tr}(\Pr_i \Lambda^{\Delta S=2}) = 0 ,\, i=1,...,4\\
\Lambda^{\Delta S=2} \equiv \Lambda^{\Delta S=2}_{0} +
\sum_{i=1}^4 Z^{\Delta S=2}_i \Lambda^{\Delta S=2}_{i}
\end{array}
\label{eq:EQD}
\end{equation}
by solving the non-homogeneous set of linear equations
\[
\sum_{j=1}^4 Z^{\Delta S=2}_j D_{ji} = -D_{0i},\, i=1,...,4
\]
The overall RC is evaluated from the condition
\begin{equation}
Z^{\Delta S=2} Z_q^{-2}{\mbox{Tr}}(\Pr_0 \Lambda^{\Delta S=2})
\vert_{\mu^2} = 1
\label{eq:NORMALIZ}
\end{equation} 
where $Z_q$ is the quark wave function RC. In~(\ref{eq:NORMALIZ}) four 
$Z_q^{-1/2}$ factors appear, because the $\Lambda$'s are four-leg 
amputated amplitudes. In Fig.~2 of~\cite{TA} the results of a number of Monte 
Carlo measurements of the 
$\<\bar K_0 \vert \widehat O^{\Delta S=2}\vert K\>$ amplitude 
are shown as a function of $m_K^2$~\cite{DS}. 
It is clearly seen that the long sought chiral behaviour of 
$\<\bar K_0 \vert\widehat O^{\Delta S=2}\vert K\>$ is correctly reproduced 
when the NP renormalization procedure
described in this section is employed, in conjunction with the
SW-Clover-Leaf improved fermionic action~\cite{SW}~\cite{HMPRS}. Similarly
good results have also been reported by the JLQCD collaboration in the case 
of the standard Wilson action~\cite{GIAP}. 

A possible difficulty with the idea of extracting the NP values of RC\&MC's 
from quark/gluon matrix elements is that a lattice gauge fixing is 
required in the simulations, leaving behind the problem of Gribov ambiguities.

The next obvious step within this approach is to go over to the much
more complicated and interesting case of the $\Delta I=1/2$ ENLWH, where the
problem of the subtraction of lower dimensional operators with power divergent
MC's has up to now forbidden a reliable computation of the 
$K\rightarrow\pi\pi\vert_{\Delta I=1/2}$ amplitude~\cite{BDHS}~\cite{GMMPP}. 
The two operators relevant in this case 
belong to the [$\bfeig$] representation of the chiral group. They 
are usually indicated by $O^{(\pm)}$ and their bare expression is 
\[
\begin{array}{l}
O^{(-)}_0 = [(\bar s \gamma^L_\mu d) (\bar u \gamma^L_\mu u)
- (\bar s  \gamma^L_\mu u) (\bar u \gamma^L_\mu d)]- \\
-[u\leftrightarrow c]\\ 

O^{(+)}_0 =  {1\over 5} [(\bar s \gamma^L_\mu d) (\bar u \gamma^L_\mu u)
+ (\bar s  \gamma^L_\mu u) (\bar u \gamma^L_\mu d)+\\
+ 2 (\bar s \gamma^L_\mu d) (\bar d \gamma^L_\mu d) + 2 (\bar s 
\gamma^L_\mu d) (\bar s \gamma^L_ \mu s)]-\\
- [(\bar s \gamma^L_\mu d) (\bar c \gamma^L_\mu c) +  (\bar s 
\gamma^L_\mu c) (\bar c \gamma^L_ \mu d)]
\end{array}
\] 
As we repeatedly said, the explicit breaking of chiral symmetry due to the
presence of the Wilson term in the action, induces the mixing of
$O^{(\pm)}_0$ with operators belonging to chiral representations other than
the $[{\bfeig}]$. Taking into account the symmetries left unbroken by the 
lattice regularization, it is easily seen that the renormalized, finite 
lattice operators, which in the fully interacting theory will transform as 
an [$\bfeig$] representation, must have the expression
\begin{equation}
\begin{array}{l}
\widehat{O}^{(\pm)} = Z^{(\pm)}[O^{(\pm)}_0 + \sum_{i=1}^4
Z^{(\pm)}_{i} O^{(\pm)}_{i} +\\
+ Z_5^{S(\pm)}\bar s \sigma_{\mu\nu} F_{\mu\nu} d 
+Z_5^{P(\pm)} \bar s \sigma_{\mu\nu} \widetilde F_{\mu\nu} d +\\+
Z_3^{S(\pm)}\bar s d  + Z_3^{P(\pm)} \bar s \gamma_5 d] 
\label{eq:Otilde}
\end{array}
\end{equation}
Let us in turn examine the various terms in~(\ref{eq:Otilde}).

$\bullet$ {\underline{Dimension 6 operators}}

The spin and color structure of the operators of dimension 6 contributing
here is obviously the same as the one of the $\Delta S=2$ case discussed
before. Only the flavor structure is different.

$\bullet$ {\underline{Dimension 5 operators}}

If the GIM mechanism is operative, the coefficients $Z_5^{S(\pm)}$ are 
actually finite, because the potential $1/a$ divergence is replaced by
a $m_c-m_u$ factor. As for $Z_5^{P(\pm)}$, GIM and CPS symmetry~\cite{BDHS} 
(CPS = CP $\times$ symmetry under $s\rightarrow d$ exchange) make it finite 
and vanishing in the limit of exact vector flavor symmetry (exact $SU(N_f)_V$).

$\bullet$ {\underline{Dimension 3 operators}} 

The GIM mechanism softens the divergence of $Z_3^{S(\pm)}$, reducing it 
from $1/a^3$ to $(m_c-m_u)/a^2$. As before, CPS symmetry makes 
$Z_3^{P(\pm)}$ vanishing in the limit of exact $SU(N_f)_V$.

An important observation at this point is that the Maiani-Testa no-go
theorem~\cite{MT} (which states that in euclidean region essentially only
matrix elements of operators between one-particle states can be extracted
from Monte Carlo data) forces us to limit our considerations to the 
$\<K \vert \widehat O^{(\pm)}\vert \pi\>$ matrix elements, leaving the
reconstruction of the $K\rightarrow\pi\pi$ amplitude to the use of the SPT's
\[
\begin{array}{l}
\<0\vert O^{\triangle S=1}_{[\bfeigs]} \vert K^0\> =
 i2\,\alpha_2\, {m^2_K - m^2_\pi\over F_\pi} \\
\<\pi^+(p)\vert O^{\triangle S=1}_{[\bfeigs]} \vert K^+(q)\> = -2\,\alpha_2\,
{m^2_K\over F_\pi^2} + 4\alpha_1\, {(p\cdot q)\over
F^2_\pi}\\ \<\pi^+\pi^-\vert O^{\triangle S=1}_{[\bfeigs]}
 \vert K^0\> = i4\,\alpha_1\, {m^2_K - m^2_\pi\over F_\pi^3}
\end{array}
\]
valid for an [\bfeig] $\Delta S=1$ chiral operator. From the above equations
it is in fact immediately seen that the $K\rightarrow\pi\pi$ amplitude can be 
obtained from the slope of $\<\pi^+(p)\vert O^{\triangle S=1}_{[\bfeigs]} 
\vert K^+(q)\>$ as a function of $(p\cdot q)$. Notice that, proceeding in this 
way, only the MC's of the parity-conserving operators contributing 
to~(\ref{eq:Otilde}) will be necessary.

Let me now briefly illustrate the strategy for the construction 
of the $\Delta I=1/2$ ENLWH (see also~\cite{TA}). 
The method we propose to arrive at a NP evaluation of the finite and 
power-divergent MC's in~(\ref{eq:Otilde}) is a generalization of the approach 
described before to deal with the case of $\widehat O^{\Delta S=2}$. 

To avoid uncontrollable numerical instabilities when the simultaneous 
computation of 
finite and infinite (in the limit $a\rightarrow 0$) MC's from a single 
(large) set of linear equations is attempted, it is more convenient
to separate the subtraction of power divergences from the subsequent finite 
mixing. We thus subdivide the whole procedure in two steps.

$\bullet$ One first defines the ``intermediate'' (power divergent) mixing 
constants, $C_5^{(\bar s d)}$ and $C_{i}^{(\bar s d)(\pm)}$, by requiring
\begin{equation}
\begin{array}{l}
\mbox{Tr}(\Pr_{\bar s d} \Lambda^{(\pm)}_{6,i}(\bar{q}q;\mu)) = 0,\ 
i=0,1,\ldots,4 \\
\mbox{Tr}(\Pr_{\bar s d} \Lambda_5(\bar{q} q;\mu))=0
\end{array}
\end{equation}
where $\Pr_{\bar s d}$ is the projector over the color and spin structure of 
the operator $\bar s d$ and the amplitudes $\Lambda^{(\pm)}_{6,i}(\bar{q}q;\mu)$
and $\Lambda_5(\bar{q} q;\mu)$ are respectively the amputated $\bar{q}, q$ 
matrix elements of the operators
\[
\begin{array}{l}
O^{(\pm)}_{6,i}= O^{(\pm)}_{i}+C_{i}^{(\bar s d)(\pm)} \bar s d ,\,\,
i=0,1,...,4 \\
O_5=\bar s \sigma_{\mu\nu} F_{\mu\nu} d +C_5^{(\bar s d)}\bar s d
\end{array}
\]

$\bullet$ The second step consists in determining the finite MC's, 
$Z^{(\pm)}_i$ and $Z^{(\pm)}_5$, by imposing the (non-homogeneous) set of 
linear conditions 
\[
\begin{array}{l}
\mbox{Tr}(\Pr_5 \widetilde{\Lambda}^{(\pm)}_{PC}(\bar{q} g q;\mu))=0 \\

\mbox{Tr}(\Pr_i \widetilde{\Lambda}^{(\pm)}_{PC}(\bar{q} q\bar{q} q;\mu))= 0,
\ i=1,\ldots,4
\end{array}
\]
where the amplitudes $\widetilde{\Lambda}^{(\pm)}_{PC}$ are the appropriately 
amputated matrix elements of the (by now only logarithmically divergent) 
parity-conserving part of the operator~(\ref{eq:Otilde}), i.e. of the operator
\[
\widetilde{O}_{PC}^{(\pm)}=\Big[O^{(\pm)}_0 + 
\sum_{i=1}^{4} Z^{(\pm)}_i  O^{(\pm)}_{6,i}\Big]_{PC}+Z^{S(\pm)}_5 O_5
\]
A feasibility study of the procedure outlined above is under way~\cite{DELTAI}.

\section{Schr\"{o}dinger Functional}

The functional integral with Dirichlet boundary conditions along 
the time direction
\beq
{\cal {K}}(\Phi_2,\Phi_1;T) = \int_{\Phi_1(\vec x)=\Phi(\vec x,0)}
^{\Phi_2(\vec x)=\Phi(\vec x,T)} 
{\cal{D}} \mu [\Phi] e^{-S[\Phi]}
\eeq
\label{eq:SF}
yields the amplitude to find the field configuration $\Phi_2=\Phi_2(\vec x)$ 
at time $t=T$, if the field configuration was $\Phi_1=\Phi_1(\vec x)$ at time 
$t=0$. It thus represents the matrix  element of the (Euclidean) transfer 
matrix, $e^{-HT}$, between the Schr\"{o}dinger states $\vert \Phi_2\>$ and 
$\vert \Phi_1\>$:
\beq
{\cal {K}}(\Phi_2,\Phi_1;T) = \<\Phi_2\vert e^{-HT}\vert \Phi_1\>
\eeq
\label{eq:TRANSFER}
More generally, given the functionals
\[
Q_k(x_k)=Q_k[\Phi(\vec{x}_k,t_k),\delta/\delta\Phi(\vec{x}_k,t_k)],\,
k=1,...,n
\]
one can compute the expectation values
\[
\begin{array}{l}
\<Q_1(x_1)Q_2(x_2)...Q_n(x_n)\>\equiv\\\\

\equiv\int_{\Phi_1}
^{\Phi_2} {\cal{D}} \mu [\Phi] e^{-S[\Phi]}
Q_1(x_1) Q_2(x_2)...Q_n(x_n)=\\\\
= \<\Phi_2\vert e^{-HT} \prod_{k=1}^n e^{Ht_k}Q_k(x_k)e^{-Ht_k}\vert \Phi_1\>
\end{array}
\]
In QCD one must take $\Phi_2=\{\vec{A}_2, \bar{\psi}_2^{(+)},\psi_2^{(-)}\}$
and $\Phi_1=\{\vec{A}_1, \bar{\psi}_1^{(-)},\psi_1^{(+)}\}$, where the 
superscripts $(\pm)$ appended to the $\bar{\psi}, \psi$ fields are there to 
remember us that only half of the fermionic components can be assigned at 
each time boundary~\cite{FAD}.

The YM and the QCD SF's have been formally constructed in the continuum,
in the temporal gauge ($A_0=0$), in~\cite{GCRT} and in~\cite{LMRY} respectively.
In~\cite{LMRY} the superscripts $(\pm)$ were taken with reference to the 
positive and negative frequency decomposition of the fermionic fields, as this 
is the most suitable choice when working in the temporal gauge. The corresponding 
constructions on the lattice, where no gauge fixing is necessary, were carried 
out in~\cite{LNWW} and in~\cite{SINT1}. In the QCD case Dirichlet boundary
conditions on the spin projections
\[
\begin{array}{l}
\psi^{(\pm)}=\frac{1\pm\gamma_0}{2}\psi,\quad
\bar{\psi}^{(\pm)}=\bar{\psi}\frac{1\pm\gamma_0}{2}
\end{array}
\]
were imposed~\cite{SINT1}. Studies of the renormalization properties of the
lattice QCD SF can be found in~\cite{SINT2} and in~\cite{LSSW}.

In absence of fermions, the SF formalism was implemented in lattice 
simulations to study the running of the gauge coupling. 
A renormalized coupling, $\alpha_s$, as a function of $q=1/L$ 
($L^d=$ lattice volume) is defined by looking at the response of $\cal K$
under small changes of an externally applied color-magnetic 
field~\cite{LSWW}. The running of $\alpha_s$ with $1/L$ is extracted in a 
recursive way, in a two step procedure~\cite{LWWP}. 

$\bullet$ Starting with a given (small) value of $g_0^2$, the variation of 
$\alpha_s$, $\Delta \alpha_s^{(1)}=\alpha_s (L)-\alpha_s (L_1=sL)$, ensuing 
from an increase of the number of lattice points by a factor $s^d$ 
($L=Na\rightarrow L_1=sNa$), is computed and its continuum limit 
is extrapolated from data taken at decreasing values of $a/L$ with fixed
$\alpha_s(L)$.

$\bullet$  Then the physical value of the lattice spacing is increased, 
at fixed renormalized coupling (by adjusting the bare coupling $g_0^2$), 
in order to have, with a number of points equal to the initial one, a 
renormalized coupling precisely equal to $\alpha_s(L_1)$, to be in 
position to start a second iteration.

At this point the whole procedure is repeated, by measuring the variation
$\Delta \alpha_s^{(2)}=\alpha_s (L_1)-\alpha_s (L_2=sL_1)$ for an increase of
the number of lattice points again by a factor $s^d$. 
At each step the limit $a/L\rightarrow 0$ of the variations 
$\Delta \alpha_s^{(i)}$ is appropriately taken.

The results obtained with this method are very accurate (few \% errors) and 
compare very nicely~\cite{ALFACOLL} with the results of~\cite{DFGP}, 
derived employing a similar iterative procedure, but using for the 
renormalized gauge coupling a definition given in terms of ratios of 
twisted Polyakov loops. For a measure of the running of 
$\alpha_s$ defined directly from the three-gluon vertex, see~\cite{HPP}.

The introduction of fermions in the SF opens the way to a wealth of very
accurate NP measurements of RC\&MC's which ultimately will allow the 
(on-shell) construction of the fully O($a$) improved QCD lattice theory. 
The use of the SF approach has the further noticeable advantages of 
providing a formalism which allows to perform Monte Carlo simulations 

$\bullet$ directly at the chiral point, $M_0=M_{cr}$, 

$\bullet$ with no need of fixing the gauge. 

Up to now the method has been used to construct (in the quenched
approximation) the O($a$) improved QCD action and the correspondingly 
improved bilinear quark operators. However, the approach seems to be 
sufficiently general to be capable to encompass the much 
more complicated and interesting case of the four-quark operators.

The general idea is to start with the SW nearest-neighbor improved fermion 
action
\begin{equation}
S_{_{SW}}=S_{_W}+a^5 \frac{i}{4}c_{_{SW}} \sum_x \bar{\psi}(x)\sigma_{\mu\nu}
P_{\mu\nu}(x)\psi(x)
\label{eq:SWACTION}
\end{equation}
where $S_{_W}$ is the standard Wilson action (from now on $r=1$) and 
$P_{\mu\nu}$ is any lattice discretization of the gauge field 
strength~\cite{PMUNU}, and to determine the value of $c_{_{SW}}$ and 
of all the necessary RC\&MC's, 
by requiring that on-shell chiral WTI's involving quark bilinear
do not have O($a$) corrections. The ``tree level'' improvement
of the Wilson action (eq.~(\ref{eq:SWACTION}) with $c_{_{SW}}=1$) kills 
in on-shell Green functions all
terms that in PT are of O($a \,(g_0^2 \log a)^n$) (i.e. the terms that in 
the continuum limit, $g_0^2\sim 1/\log a$, are effectively of order $a$),
leaving uncancelled O($a$) subleading logarithmic corrections~\cite{HMPRS}.

To illustrate the SF method let me describe the strategy for the determination
of quantities like $c_{_{SW}}$, $Z_A$, $Z_V$ or the critical mass, $M_{cr}$, 
which are only functions of $g_0^2$. More subtle is the problem of 
computing the renormalized coupling constant, $g_R^2$, or the renormalized 
quark mass, $m_R$, or certain MC's, because it turns out that the general 
mass-independent renormalization scheme, consistent with O($a$) improvement, 
requires a rescaling of the bare parameters by mass dependent 
factors~\cite{LSSW}.

The crucial observation which makes the whole program actually feasible is
that, for the sake of computing the functional dependence of $c_{_{SW}}$, 
$Z_A$, $Z_V$, or $M_{cr}$, on $g_0^2$, for small $g_0^2$, all the necessary 
numerical simulations can be quite happily performed deep in the perturbative 
region. This can be realized by working in a not too large volume in order 
to keep lattice momenta, $p_n=\frac{2\pi}{L}n$, much larger than the typical 
mass scale of the theory. Indicatively, if $N$ is the number of points per 
lattice side, one should have
\[
2\pi/L\gg\Lambda_{QCD}\,\, \mbox{i.e.}\,\, N \ll \exp(1/2b_0g_0^2)
\]
At the same time to keep under control discretization effects, which scale
like powers of $a/L$ (times possible logarithmic factors), one must also require
\[
L/a=N\sim\mbox{large}
\]
With the choices $g_0^2 < 1$ and $N\gtapprox 10$ both conditions are well 
satisfied and the physical volume of the box (determined, for instance, as 
explained in~\cite{LSWW}) turns out to be rather small ($L\simeq 0.5$ fm). An 
important consequence of these choices is that simulations can be performed 
directly at the critical mass, $M_0=M_{cr}$, as for small $g_0^2$ the 
lowest eigenvalue of the Dirac operator is O($1/L$) and not O($M_0-M_{cr}$).

\begin{figure}[htb]
\vspace{6cm}
\includegraphics{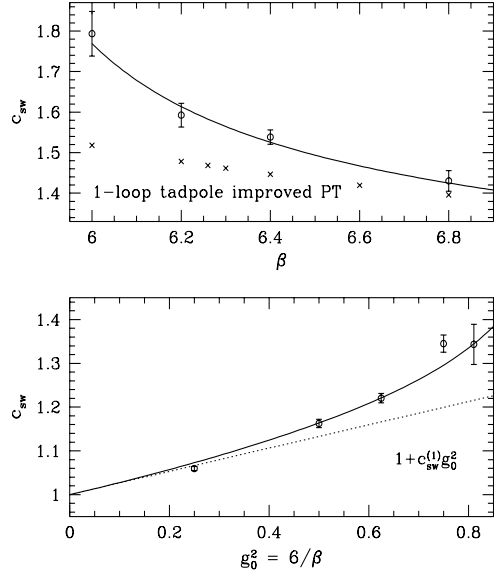}
\null\vskip 0.5cm
\caption{$c_{_{SW}}$ for large and small $g_0^2$. Comparison with 1-loop TIPT 
(crosses) and PT (dots) is shown. The solid line is a numerical interpolation 
of the data.}
\label{fig:csw}
\end{figure}


To illustrate the method in a concrete way, let me discuss the determination
of $c_{_{SW}}(g_0^2)$ and $M_{cr}(g_0^2)$. One starts from the lattice PCAC
equation
\begin{equation}
\<\partial_\mu A_\mu^f (\vec{x},x_0) O^f\>=
2m\<P^f(\vec{x},x_0)O^f\>+\mbox{O}(a)
\label{eq:PCAC}
\end{equation}
where ($x\notin \mbox{Support of}$ $O^f$)
\[
A_\mu^f=\bar{\psi}\gamma_\mu\gamma_5\frac{\tau^f}{2}\psi, \,\,
P^f=\bar{\psi}\gamma_5\frac{\tau^f}{2}\psi
\]
$O^f$ is an operator which creates a pseudoscalar state at $t=0$. Explicitly
one can take
\[
O^f=-\int_0^Ld^3y\int_0^Ld^3z\frac{\delta}{\delta\psi_1^{(+)}(\vec{y})}
\gamma_5\frac{\tau^f}{2}\frac{\delta}{\delta\bar{\psi}_1^{(-)}(\vec{z})}
\]
In~(\ref{eq:PCAC}) it is understood that the fermionic boundary values are set
to zero after the action of the grassmannian functional derivatives.
A plot of 
\[
R_A=\<\partial_\mu A_\mu^f (\vec{x},x_0) O^f\>/
\<P^f(\vec{x},x_0)O^f\>
\]
as a function of the current insertion time, $x_0/a$, shows strong violations
of chirality~\cite{JLLSSSWW}, when the standard Wilson action is employed.
These effects are to be attributed to the (large) O($a$) corrections affecting
eq.~(\ref{eq:PCAC}). They can be eliminted from all Green 
functions at unequal points, following the steps described below

$\bullet$ add the SW-term, $\frac{i}{4}c_{_{SW}} a^5\sum_x \bar{\psi} 
\sigma_{\mu\nu} P_{\mu\nu}\psi$, to the standard Wilson action,

$\bullet$ construct renormalized improved quark operators, $\widehat{O}^I_q$, 
by adding to their bare expression, $O_q$, a suitable linear combination of 
operators, $O^{(i)}_q$, with dimensions up to $\mbox{dim} O_q^{(i)} = 
\mbox{dim} O_q + 1$,

$\bullet$ fix, at any given value of $g_0^2$, $c_{_{SW}}$ and the corresponding
MC's by requiring O($a$) corrections to be absent from WTI's.

The renormalized, O($a$)-improved expressions of the axial current
and of the pseudoscalar density are ($m_q\equiv M_0-M_{cr}$)
\begin{equation}
\begin{array}{l}
\widehat{A}_\mu^{fI}=Z_A[(1+a m_q b_A)A_\mu^f + a c_A \partial_\mu P^f]\\
\widehat{P}^{fI}=Z_P(1+a m_q b_P)P^f
\end{array}
\end{equation}
where at tree level, beside $c_{_{SW}}^{tree}=1$, one has 
$b_{A}^{tree}=b_{P}^{tree}=1$, $c_{A}^{tree}=0$~\cite{HMPRS}. 
At any given $g_0^2$ the improved ratio
\[
\widehat{R}^I_A=\<\partial_\mu \widehat{A}_\mu^{aI} (\vec{x},x_0) O^f\>/
\<\widehat{P}^{aI}(\vec{x},x_0)O^f\>
\]
is measured for different values of $x_0/a$ and for two different (zero and 
non-zero) boundary gauge field configurations. Actually to the order one is 
working (terms O($a^2$) are neglected), the behavior of $\widehat{R}^I_A$ as 
a function of $x_0/a$ is
only sensitive to $c_{_{SW}}$ and $c_{A}$. In fact one can write
\[
\begin{array}{l}
\widehat{R}^I_A=\frac{Z_A}{Z_P}\frac{1+am_qb_A}{1+am_qb_P}
\frac{\<(\partial_\mu A_\mu^f + ac_A\partial^2 P^f) O^f\>}
{\<P^f O^f\>} + \mbox{O}(a^2)
\end{array}
\]
The nice results obtained for $c_{_{SW}}(g_0^2)$ and $c_{A}(g_0^2)$  
by imposing that $\widehat{R}^I_A$ is independent of i) $x_0/a$, ii) the 
particular form of the operator $O^f$ iii) the boundary values of the gauge 
fields, are shown in Figs.~1 and~2. While $c_A$ is negligible to 
all practical extents, $c_{_{SW}}$ appears to be a rapidly increasing 
function of $g_0^2$. Data for $c_{_{SW}}$ are larger than expected from simple 
``tadpole improvement'' arguments~\cite{LPMK}.
\begin{figure}[htb]
\vspace{4.5cm}
\includegraphics{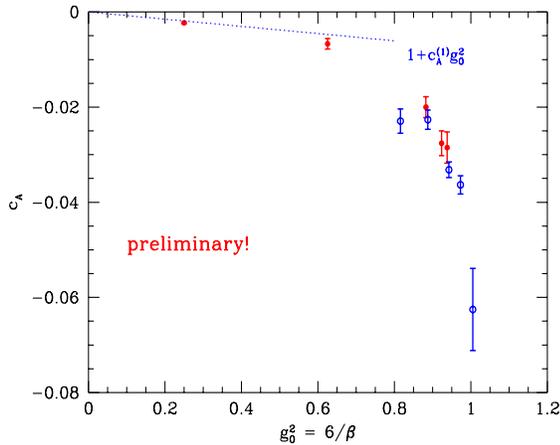}
\null\vskip 0.1cm
\caption{$c_{A}$ as a function of $g_0^2$. The dotted line is 1-loop PT.}
\label{fig:ca}
\end{figure}
Once $c_{_{SW}}$ and $c_{A}$ have been fixed, by varying $M_0$, one can
determine $M_{cr}$ as the value of $M_0$ for which $R_A$ (and of course 
also $\widehat{R}^I_A$) vanishes. Data for $\kappa_{cr}$, measured at large 
and small $g_0^2$, are shown in Fig.~\ref{fig:kappa}, 
together with the results from 1-loop Tadpole Improved PT (TIPT) and from 
1-loop straight PT, respectively. 
\begin{figure}[htb]
\vspace{5.1cm}
\includegraphics{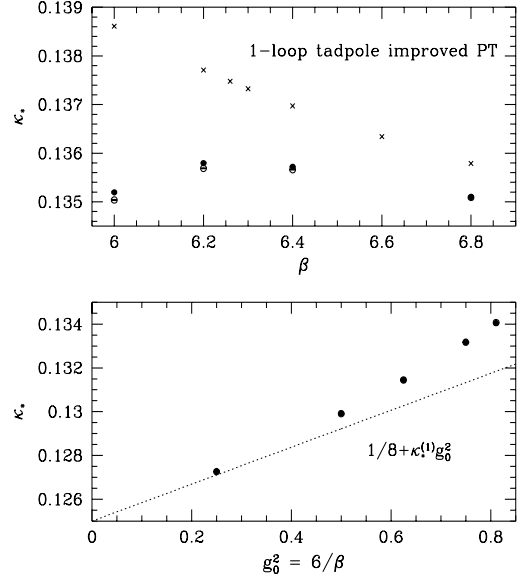}
\null\vskip 0.8cm
\caption{$\kappa_{cr}$ for large and small $g_0^2$. Comparison 
with 1-loop TIPT (crosses) and PT (dots) is shown.}
\label{fig:kappa}
\end{figure}
It is clearly seen that the non-monotonic behavior of 
$\kappa_{cr}$ is in striking contrast with tadpole improvement 
expectations. This is not surprising in view of the fact that a proper 
definition of $M_{cr}$ requires a linearly divergent subtraction.
\begin{figure}[htb]
\vspace{4.2cm}
\includegraphics{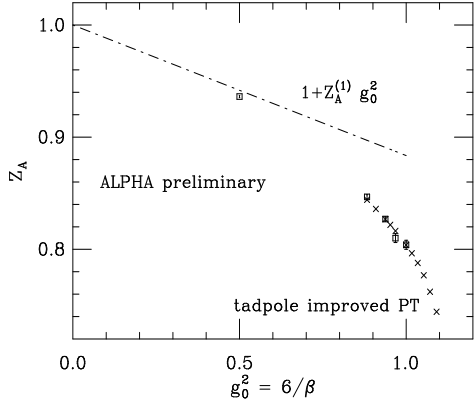}
\null\vskip 0.0cm
\caption{$Z_A$ as a function of $g_0^2$. Comparison with 1-loop
TIPT (crosses) and PT (dots) is shown.}
\label{fig:za}
\end{figure}
Absence of O($a$) corrections in WTI's is a strong requirement which also 
allows an accurate evaluation of the current RC's. In Figs.~\ref{fig:za} 
and~\ref{fig:zv} we show 
the results for $Z_A$ and $Z_V$, as functions of $g_0^2$, together with the 
predictions of 1-loop TIPT and PT. In both cases numbers from 
TIPT interpolate 
quite nicely Monte Carlo results. Perhaps the lesson we can draw from the 
different ability of TIPT in reproducing data directly extracted from 
simulations is that TIPT can be really useful only for estimating 
finite RC\&MC's.

We would like to conclude this section by stressing again that the use of the SF
formalism in Monte Carlo simulations seems to be extremely promising and has already
provided us with a lot of very accurate results. It would really be a great
achievement if one could extend this approach to the construction of finite,
renormalized four-quark operators. 

\section{NP renormalization by ``heating - cooling'' steps}

For completeness, in this section I wish to discuss a completely different NP
renormalization method based on ``heating-cooling'' Monte Carlo steps. The
method is designed to deal with the problem of giving finite and 
renormalized field theoretical definitions of the topological charge density, 
$q(x)$ and of the topological susceptibility, 
$\chi$, in lattice gauge theories~\cite{diG1}. The method is not new, 
but is has been recently tested with success in certain exactly solvable 
$\sigma$-models~\cite{ABF} 
and, more interesting, it has revealed an unexpected vanishing of $\chi$ 
across the YM deconfining temperature~\cite{ADEdiG}.

\begin{figure}[htb]
\vspace{4.2cm}
\includegraphics{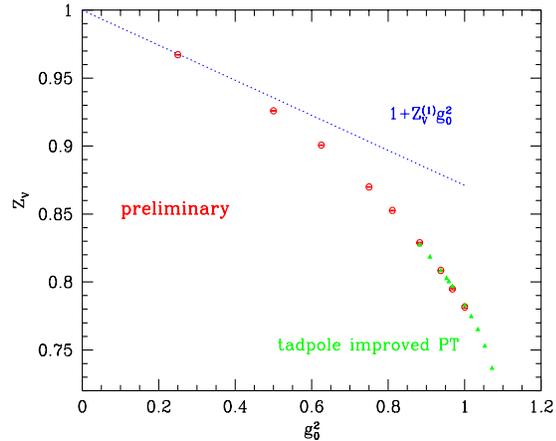}
\null\vskip 0.1cm
\caption{$Z_V$ as a function $g_0^2$. Comparison with 1-loop TIPT (triangles) 
and PT (dots) is shown.}
\label{fig:zv}
\end{figure}

The topological susceptibility has been the object of many 
theoretical and numerical investigations not only because it is directly 
related to the  mass of the flavor singlet pseudoscalar meson 
($m^2_{\eta_{S}}=\frac{2 N_f}{F^2_\pi}\chi$)~\cite{WV}, but also 
because it is a natural question to ask whether and how it is possible to give 
on a lattice a notion which would reduce to the standard notion of topology in
the continuum limit. For lack of space (see however~\cite{GP} and references
therein), I will not mention definitions of the
topological charge, based on purely geometrical considerations (the first
of which dates back to~\cite{L}), nor the construction of $\chi$, based on
the use of the flavor singlet axial WTI~\cite{BRTY}. These two topics are,
anyway, outside the aim of this talk, as in both cases, by construction, no
renormalization is expected to be necessary.

The formula for the topological charge in a continuum 
gauge theory (Pontrjagin number) is $Q=\int d^4x q_c(x)$,
where $q_c(x)=\frac{g_0^2}{32\pi^2}\epsilon_{\mu\nu\rho\sigma}
\mbox{Tr}(F_{\mu\nu}F_{\rho\sigma})$ 
is the topological charge density. On a lattice one can use the definition
\[
q(x)=-\frac{1}{2^4 32\pi^2}\sum_{\mu\nu\rho\sigma=\pm 1}^{\pm 4}
\epsilon_{\mu\nu\rho\sigma}\mbox{Tr}(P_{\mu\nu} P_{\rho\sigma})
\]
where $P_{\mu\nu}$ is a discretization of $F_{\mu\nu}$~\cite{PMUNU}. 
It is, in fact, immediately seen that, in the naive $a\rightarrow 0$ limit, 
$q(x)\rightarrow a^4 q_c(x) + \mbox{O}(a^6)$. This means that in the continuum
(field-theoretical) limit, $g_0^2\rightarrow 0$ with $g_0^2\sim 1/\log a$,
we will have
\begin{equation}
q(x)\rightarrow Z_Q(g_0^2) a^4(g_0^2)q_c(x) + \mbox{O}(a^6)
\label{eq:TOPCH}
\end{equation}
From the previous equations it naturally follows that the lattice version of
the continuum topological susceptibility, 
$\chi_c=\int d^4x \<T(q_c(x) q_c(0))\>$, can be taken to be
\[
\chi=\sum_x \<T(q(x) q(0))\>
\]
According to the general rules of field theory, $\chi_c$ and $\chi$ will be
related in the continuum limit by
\begin{equation}
\chi= Z_Q^2(g_0^2) a^4(g_0^2) \chi_c + C_{\chi}(g_0^2) + \mbox{O}(a^6)
\label{eq:TOPSUS}
\end{equation}
The factor $Z_Q^2(g_0^2)$ comes from~(\ref{eq:TOPCH}), while the subtraction
term, $C_{\chi}(g_0^2)$, is a consequence of the mixing of $\chi$ with the 
operator $\mbox{Tr}(F_{\mu\nu}F_{\mu\nu})$ and the identity. A NP estimate 
of $Z_Q(g_0^2)$ and $C_{\chi}(g_0^2)$ relies on the simple observation that
short range fluctuations (at the scale of the cut-off) are responsible for
renormalization effects, while physical effects, like confinement, come from
much larger distances (of the order of the correlation length, $\xi$). When 
approaching the continuum limit, the two scales are widely separated. Using 
a local Monte Carlo updating algorithm, fluctuations at distances $\sim a$ 
will be soon thermalized, whereas fluctuations at the scale of $\xi$ are 
critically slowed down. 
\begin{figure}[htb]
\vspace{4.2cm}
\includegraphics{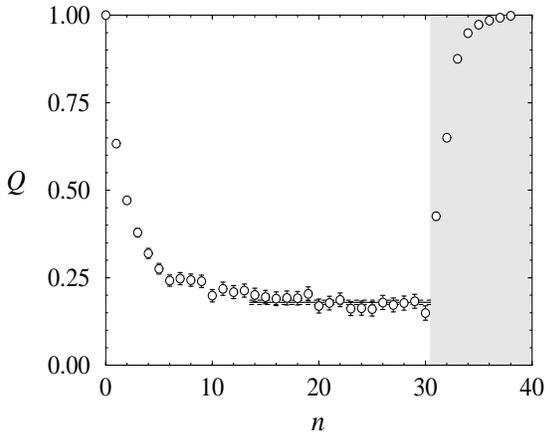}
\null\vskip 0.0cm
\caption{$Q$ as a function of the heating step, $n$.}
\label{fig:Q}
\end{figure}
For a standard local algorithm, like Metropolis, the time (\# of sweeps)
necessary to thermalize fluctuations at distances $d$ grow proportionally
to $d^z$ with $z\sim 2$. Changes in the global (topological) properties of the 
gauge configuration are expected to require a much longer (exponential) time.
To measure $Z_Q$ a gauge configuration with topological charge 1 is
placed on the lattice and, at any given temperature, $\beta=2 N_c/g_0^2$,
it is progressively heated (only link-variable changes that make the action
increase are accepted) for a few Monte Carlo sweeps. $Z_Q$ is the value 
at which $Q$, measured as a function of the Monte Carlo step, $n$, reaches 
a plateau (Fig.~\ref{fig:Q}). Data have 
very small errors and $Z_Q$ can be extracted with remarkable accuracy. The last 
points of Fig.~\ref{fig:Q} (shaded area) are obtained by cooling back the 
configuration. It is seen that the initial value $Q=1$ is reobtained, thus 
showing that the global properties of the successive gauge configurations one 
has gone through were left unchanged by the heating process.
Once $Z_Q$ is known, $C_{\chi}$ is determined starting from a 
(topologically) trivial gauge configuration and measuring $\chi$ for a few
Monte Carlo steps, until it reaches a plateau (Fig.~\ref{fig:chi}). Since 
in~(\ref{eq:TOPSUS}) the first term is zero ($Q=0$), the height of the plateau 
is precisely $C_{\chi}$. Cooling back the gauge configuration, one checks that 
no unwanted  non-trivial contributions to $\chi$ have been introduced during 
the heating process.
\begin{figure}[htb]
\vspace{4.1cm}
\includegraphics{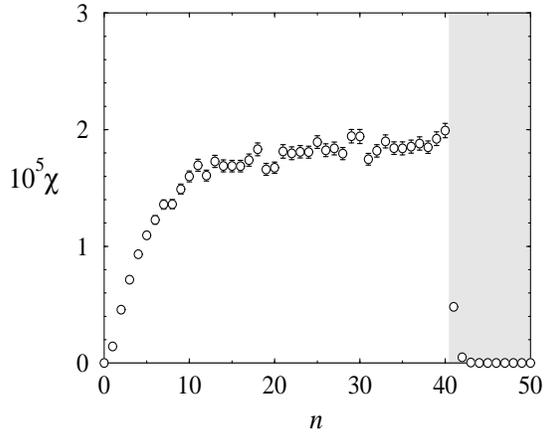}
\null\vskip 0.1cm
\caption{$\chi$ as a function of the heating step, $n$.}
\label{fig:chi}
\end{figure}

\section{Short distance expansions, higher twists and condensates}

Short Distance Expansions (SDE's) are among the very few approaches, besides 
instanton calculus and SPT's, that may provide pathways to NP analytical 
calculations in (non-supersymmetric) QCD. It is then of the utmost importance 
to understand to what extent corrections to the leading term of the expansion 
can be rigorously defined and reliably computed. The heart of 
the problem is that, while the calculation of Wilson coefficients is 
necessarily truncated to some finite order in $\alpha_s$, exponentially small 
($\sim \exp (-c/\alpha_s)$) non-perturbative corrections (as those given by 
next order terms) are retained.


The whole issue is complicated by the difficulties arising from the 
existence of renormalon ambiguities in the Borel resummation of PT. As is well 
known~\cite{tHL}, perturbative series in field theories are only asymptotic
and in the most interesting cases (gauge theories) are not Borel summable,
as a consequence of the presence of singularities located on the positive real 
axis of the Borel $u$-plane. We recall that a pole at the position $u=u_0>0$
leads to an ambiguity of the kind $\exp(-\frac{c}{g_0^2}u_0)$
in the Borel resummed series.

In all instances where a large scale expansion is performed 
($e^+ e^-$-annihilation, DIS scattering, Heavy Quark Effective 
Theory (HQET),...), further renormalon singularities appear. As in the full
theory, they come from the factorial growth of the weight of
certain classes of diagrams when the running coupling $g^2(k)$ ($k$ is the 
loop momentum) is expanded as a power series in $g^2(\mu)$ ($\mu$ is some 
fixed renormalization scale) \footnote{There exist also Borel 
singularities coming from the factorial growth of the number of diagrams 
with the order of PT. They lie on the negative side of the real axis
(think of the typical case of $\lambda \phi^4$) and do not affect Borel 
summability properties. They are not of interest in this discussion.}.
The situation can be summarized as follows.

$\bullet$ Borel poles coming from the loop integration region where 
$-b_0\log (k/\mu) > 0$, i.e. where $k < \mu$ ($b_0 > 0$ in our conventions), 
are called Infra-Red (IR) renormalons and affect Wilson coefficients.

$\bullet$ Poles coming from the loop integration region where 
$-b_0\log (k/\mu) < 0$, i.e. where $k > \mu$, are called Ultra-Violet (UV) 
renormalons. They show up in the hadronic matrix elements of the 
Wilson operators.

A lot of interesting work has been recently done (for a nice review 
see~\cite{S}) on the question of the appearance and cancellation of 
renormalon ambiguities. By matching the Wilson expansion against full 
theory calculations, it was shown that SDE-renormalon 
ambiguities must cancel between Wilson coefficients and hadronic matrix 
elements. The way it happens is rather involved and the details depend 
on whether one uses a soft (e.g. dimensional) or a hard (e.g. lattice) 
regularization.


In spite of this nice result, it has been argued in~\cite{MS} that a further
strong assumption on the relative magnitude of perturbative corrections 
{\it vs} NP terms is necessary and unavoidable, if one wants to give a precise 
meaning to large scale expansions beyond the leading term. 
To illustrate the point I will closely follow ref.~\cite{MS}.

Let ${\cal{P}}_{fi}(Q^2)\equiv\<f\vert\widehat{P}(Q^2)\vert i\>$ be the Fourier 
transform of some bilocal T-product of operators. The expansion of 
${\cal{P}}_{fi}(Q^2)$ for large $Q^2$ has the form
\begin{equation}
\begin{array}{l}
{\cal{P}}_{fi}(Q^2)=C_1(\frac{Q^2}{\mu^2})\<f\vert O_1(\mu)\vert i\> +\\ 
+ \frac{1}{(Q^2)^n} \, C_2(\frac{Q^2}{\mu^2})\<f\vert O_2(\mu)\vert i\>
\label{eq:EXPANSION}
\end{array}
\end{equation}
where $O_1(\mu)$ and $O_2(\mu)$ are local operators normalized at scale $\mu$, 
with $\mbox{dim} O_2=\mbox{dim} O_1 + 2 n$. For simplicity only two terms 
have been included in~(\ref{eq:EXPANSION}), but the argument can be readily 
extended to any number of terms. The operator $\widehat{P}$ is renormalized
(in the full theory) at a scale $M$ and the dependence of $C_1$ and $C_2$
on $M$ is understood. The usefulness of the expansion~(\ref{eq:EXPANSION})
lies in the fact that short distance NP effects are contained in the matrix
elements of the operators $O_i$, while the $C_i$ can be reliably computed 
in PT at large $\mu$.
It is assumed that the matrix element of $O_1$ is exactly known, either 
because it is the identity or because it is a conserved operator. The question 
is whether one can unambiguously define $O_2$ by matching the large $Q^2$
(perturbative) computation of the rhs of~(\ref{eq:EXPANSION}) in the full 
theory with the form of the expansion in the lhs. Barring exceptional cases, 
the interesting situation is when $O_2$ mixes with $O_1$.

If dimensional regularization is used, IR renormalon singularities will
appear in the Borel transform of $C_1$, making its perturbative series 
non-Borel summable. Matching implies that a compensating UV renormalon 
ambiguity will affect the matrix elements of $O_2$. This simple argument 
already shows where the heart of the problem with the definition of higher 
twist operators (such as $O_2$) lies: the perturbative series for the Wilson 
coefficients have renormalon ambiguities that are of the same (exponentially
small) magnitude of the NP effects described by the higher dimensional terms
one is including in the Wilson expansion.

One might imagine to overcome this difficulty by alternatively trying to
define higher order terms in SDE's by either comparing different expansions
involving the same higher dimensional operators (at the expenses of a certain
number of predictions) or by resorting to a NP, say lattice, determination of
the matrix elements of the relevant higher twist operators. It turns 
out~\cite{MS} that the two methods are equivalent and offer only 
a partial solution of the problem, in the sense that 

1) - after renormalizing $O_2$, $C_1$ is free of the leading renormalon 
ambiguity of order $\frac{1}{(Q^2)^n} \sim \exp (-\frac{4\pi}{b_0 
\alpha_s(Q^2)} 2 n)$. 

2) - Extra renormalon poles, located further away along the positive real Borel
axis are assumed to be related to the presence of even higher dimensional 
operators in~(\ref{eq:EXPANSION}) and could be in turn eliminated by 
sacrificing extra physical predictions to fix their ambiguous matrix elements.

3) - However, since the Wilson coefficients can only be known up to a 
certain order in PT (say, up to order $(\alpha_s)^k$) and the cancellation
of the leading renormalon starts to be numerically effective at some
large order in PT (the higher the dimension of the involved operators, 
the larger the order the cancellations start to be operative), we must 
require higher order perturbative contributions to be negligible with respect 
to the exponentially small terms we are retaining in the expansion.

In formulae this means that for the expansion~(\ref{eq:EXPANSION}) to
be useful the inequality
\begin{equation}
\begin{array}{l}
(\alpha_s(Q^2))^{k+1}\ltapprox\exp[-\frac{4\pi|\mbox{dim}O_2-\mbox{dim}O_1|}
{b_0 \alpha_s(Q^2)}]
\label{eq:INEQ}
\end{array}
\end{equation}
must hold \footnote{Technically the cancellation of the leading 
renormalon ambiguity in $C_1$ ensures that the factorially growing 
perturbative tail one is neglecting does not sum 
up to an exponential factor larger than the one 
appearing in~(\ref{eq:INEQ}).}. Unfortunately, whether or not inequalities 
like~(\ref{eq:INEQ}) are numerically satisfied in real life cannot be decided 
{\it a priori}.

The situation is even more troublesome in the case of ``condensates'', i.e.
when one is dealing with the vacuum expectation value (v.e.v.) of the 
T-product of, say, two currents, like in $e^+ e^-$-annihilation. The v.e.v's of
the local operators appearing in the SDE are the NP quantities one would like
to define properly and utilize to make predictions in other physical processes.
Naively, i.e. according to what we were taught in our graduate courses
in field theory, when a local operator, $O$, can mix with the identity, its 
v.e.v. will be power divergent and should be subtracted out. Furthermore, 
whatever the subtraction prescription is, physical results should 
not depend on it. Thus without loss of generality, one can always use the 
prescription $\widetilde{O} = O-\<O\>\id$. The consequence of this argument
is that the v.e.v. of a local operator cannot have a physical meaning.
The only exception is when the v.e.v. in question plays the role of order 
parameter of some symmetry. The most obvious example of this situation is  
$\<\bar{\psi} \psi\>$, which is the order parameter of chiral symmetry. 
In this case one can use the WTI of the chiral symmetry to define what is to 
be meant by $\<\bar{\psi} \psi\>$ and, if required, to relate its value
to the mass of the Goldstone boson, when $m_q\neq 0$. 

This situation should be contrasted with the case of the condensate
$F^2\equiv\<\mbox{Tr}(F_{\mu\nu}F_{\mu\nu})\>$. $F^2$ is not the order 
parameter of any symmetry and, in fact, it does not appear in any useful
WTI. Besides, if we recall that the v.e.v. of the trace of the energy-momentum 
density tensor, $\<\theta_{\mu\mu}\>$, is proportional to $F^2$, we would be
very strongly tempted to conclude that $F^2=0$, to prevent the vacuum to have
infinite total energy. The last statement follows from the observation
that Lorentz invariance implies $\<\theta_{\mu\nu}\>= k g_{\mu\nu}$, so that 
vacuum energy finiteness requires $\<\theta_{00}\>=0$, leading 
to the conclusion $k=0$ and hence $F^2=0$.


{\bf Acknowledgments}

I would like to thank all my collaborators for innumerable conversations. My
special thanks go to S. Capitani, M. L\"{u}scher, G. Martinelli, A. Pelissetto, 
R. Petronzio, C. T. Sachrajda, R. Sommer, M. Talevi, M. Testa, A. Vladikas.
I should also thank R. Sommer and H. Wittig for sending me Figs. 1 to 5 prior
to publication.

\end{document}